\documentclass[12pt]{article}

\usepackage{latexsym,amssymb}
\usepackage{graphics}
\usepackage{graphicx}
\usepackage{epsfig}

\begin{document}

\title{Quantum Scaling Approach to Nonequilibrium Models}

\author{T. Hanney\thanks{Present address: School of Physics,
University of Edinburgh, Mayfield Road, Edinburgh, EH9 3JZ, UK} and
R.B. Stinchcombe
\\ {\small Theoretical Physics, 1 Keble Road, Oxford, OX1 3NP, UK}}

\maketitle

\begin{abstract}
Stochastic nonequilibrium exclusion models are treated using a real
space scaling approach. The method exploits the mapping between
nonequilibrium and quantum systems, and it is developed to accommodate
conservation laws and duality symmetries, yielding
exact fixed points for a variety of exclusion models. In addition, it is
shown how the asymmetric simple exclusion process in one dimension can
be written in terms of a classical Hamiltonian in two dimensions using
a Suzuki-Trotter decomposition. 
\end{abstract}


\section{Introduction}
 
Stochastic models of lattice gas dynamics provide insight into the
nonequilibrium behaviour of a variety of physical processes, such as
surface reactions and growth, catalysis and transport phenomena. These
models are systems of many interacting particles --- the dynamics of
the particles are prescribed in the model definition --- and the
evolution is typically governed by a Master equation. They exhibit
steady state phase transitions and very rich dynamics, but there
exists no general framework in which to analyse nonequilibrium models. 
Exact treatments are scarce (see
e.g. \cite{alcaraz94,grynberg95,derrida93})
and so approximate techniques are required. To this end we present a
scaling treatment designed to capture universal and
non-universal critical properties of nonequilibrium systems.

The scaling method was developed in detail in \cite{stella83}. It
exploits the well known equivalence between the 
Master equation and the Schr\"odinger equation in imaginary time
\cite{alcaraz94} --- the stochastic lattice gas model is written as a quantum
spin model. The scaling is achieved using a real space blocking
procedure \cite{neimeijer76} in order to thin out the number of
degrees of freedom. It has been applied to the contact process
\cite{hooyberghs,hooyberghs01}, where very accurate results for the
critical point and certain critical exponents were obtained. Here, we
show how to adapt the method to models which possess a conservation
law (e.g. conserved particle number) in order to obtain exact fixed
points. Further, from a stability analysis of the fixed points, we
infer the role of bias in the model dynamics.    


\section{Quantum Scaling for Exclusion Models}

In the following, we consider exclusion models --- models where sites on a
lattice are either occupied by a single particle or vacant. In the
quantum formulation, these models are spin-1/2 quantum chains. The
mapping is achieved  
by interpreting configurations of particles and vacancies in
the nonequilibrium model as a configuration of quantum spins, where
particles are replaced by an up-spin and vacancies are replaced by a
down-spin, say. Since the dynamics in the nonequilibrium model become
processes involving spin flips, they can be expressed in terms of a
quantum Hamiltonian. Hence, for a lattice containing $L$ sites, the
configuration is written $\vert \{\sigma_l\} \rangle = \prod_{l=1}^L
\otimes \vert \sigma_l \rangle$, where
$\{\sigma_l\}=\sigma_1,\ldots,\sigma_L$. We use the notation
$\sigma_l = +_l$ to represent an up-spin at site $l$ (i.e. a particle in
the nonequilibrium system), and $\sigma_l = -_l$ to represent a
down-spin (i.e. a vacancy in the nonequilibrium system).
The steady state of the nonequilibrium system is equivalent to the
ground state of the corresponding quantum problem.
     

\subsection{Quantum Renormalisation Group Scheme}

\begin{figure}
\begin{center}
\unitlength=1mm
\linethickness{0.4pt}
\begin{picture}(65,40)
\put(0,33){\line(1,0){65}}
\put(10,32){\line(0,1){2}}
\put(25,32){\line(0,1){2}}
\put(40,32){\line(0,1){2}}
\put(55,32){\line(0,1){2}}
\put(8,32){$\underbrace{\phantom{\line(1,0){19}}}$}
\put(38,32){$\underbrace{\phantom{\line(1,0){19}}}$}

\put(32.5,10){\line(0,1){16}}
\put(31.65,10.5){$\downarrow$}
\put(33,17){Blocking transformation}

\put(0,3){\line(1,0){65}}
\put(17.5,2){\line(0,1){2}}
\put(47.5,2){\line(0,1){2}}

\put(9,34.5){1}
\put(24,34.5){2}
\put(39,34.5){1}
\put(54,34.5){2}
\put(16.5,5){$\nu$}
\put(43.6,5){$\nu+1$}
\put(16,37){$\nu$}
\put(43.6,37){$\nu+1$}

\end{picture}
\caption{Blocking of the lattice for $b=2$.} \label{fig:blocking}
\end{center}
\end{figure}
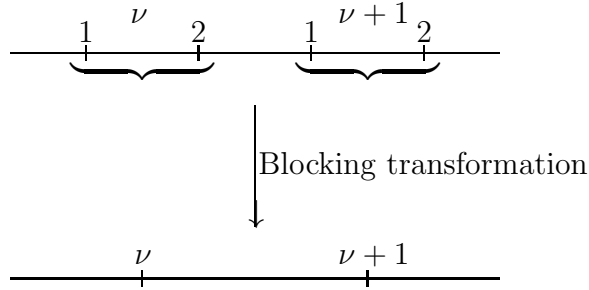

We now outline the renormalisation group scheme for quantum systems
\cite{stella83}. The first step 
is to divide the lattice into adjacent blocks, each containing $b$
sites, as indicated in figure \ref{fig:blocking}. 
The lattice contains $L' = L/b$ blocks, labelled $\nu=1, \dots, L'$,
which will form the sites in the renormalised system. 
The Hamiltonian for the blocked system is written in terms of an intrablock
part $H_{\nu}$, containing all the interactions within block $\nu$, and
an interblock part $H_{\nu,\nu+1}$, containing all the interactions
between block $\nu$ and block $\nu+1$ (we assume that there are
nearest neighbour interactions only). Hence
\begin{equation}
H = \sum_{\nu=1}^{L'} \left[ H_\nu + H_{\nu,\nu+1} \right]\;.
\end{equation}
The renormalisation is achieved as follows. We treat the intrablock
Hamiltonian $H_{\nu}$ exactly, and regard the interblock part
$H_{\nu,\nu+1}$ as a perturbation. Thus, we find the eigenvectors of
$H_{\nu}$ and use only those of
lowest energy to form a truncated basis of states --- in a spin-1/2
system we aim to keep the two lowest lying eigenstates. 
The configuration of the renormalised lattice is written as a direct product
over blocks:
\begin{equation}
\vert \{\sigma_{\nu}\} \rangle = \prod_{\nu=1}^{L'} \otimes \vert
\sigma_{\nu} \rangle\;,
\end{equation}
where $\{\sigma_{\nu}\} = \sigma_1, \ldots, \sigma_{L'}$ and the block
spin states $\vert \sigma_{\nu} \rangle$ are the renormalised 
basis of states obtained from the lowest lying eigenstates of
$H_\nu$. Then the renormalised Hamiltonian $H'$ is obtained by
writing $H' = \sum_\nu [ H'_\nu+ H'_{\nu,\nu+1} ]$. The matrix
elements of $H'_\nu$ are given by
\begin{equation} 
\langle \sigma_\nu | H'_\nu | \sigma'_\nu \rangle
=\langle \sigma_\nu | H_\nu | \sigma'_\nu \rangle\;,
\end{equation}
where on the r.h.s., we use the fact that we have assigned
eigenvectors of $H_\nu$ to the block spin states (since these
eigenvectors are orthogonal, i.e.\ $\langle \sigma | \sigma' \rangle =
\delta_{\sigma, \sigma'}$ although $H_\nu$ is not Hermitian, $H'_\nu$
is diagonal). Similarly, the matrix elements of $H'_{\nu,\nu+1}$ are
given by
\begin{equation} \label{V}
\langle \sigma_\nu |H'_{\nu,\nu+1} | \sigma'_\nu \rangle =
  \langle \sigma_\nu | H_{\nu,\nu+1} | \sigma'_\nu \rangle \;, 
\end{equation}
and thus $H'_{\nu,\nu+1}$ contributes the interaction terms in the
renormalised Hamiltonian $H'$.

Thus we have a prescription whereby we retain only the lowest lying
eigenstates of a block Hamiltonian in order to thin out the number of
degrees of freedom and still retain the features important for
criticality --- the renormalisation is carried out near the ground
state of the quantum system, equivalent to the steady state of the
nonequilibrium system. By assigning this truncated basis of states to
a block spin variable in an appropriate way, one hopes to obtain a
rescaled Hamiltonian $H'$ of the same form as the original Hamiltonian
$H$ but with rescaled parameters.


\subsection{Scaling for the Asymmetric Simple Exclusion Process}
\label{sect:quant_ASEP}

In this section, we apply the above transformation to the asymmetric
simple exclusion process (ASEP) in one dimension with periodic
boundary conditions. In the ASEP, particles 
hop to the right (left) with rate $p$ ($q$) provided that the target
site is empty. These dynamics can be expressed in terms of a quantum
Hamiltonian given by
\begin{equation} \label{quant_ASEP}
H = \sum_{l=1}^L \left[
p\:(P_{l}^{+}P_{l+1}^{-}-\sigma_{l}^{-}\sigma_{l+1}^{+})
+q\:(P_{l}^{-}P_{l+1}^{+}-\sigma_{l}^{+}\sigma_{l+1}^{-}) \right] .   
\end{equation}
where  $P_{l}^{\pm} = \frac{1}{2}(1\pm\sigma_{l}^{z})$ are projection 
operators and $\sigma_{l}^{+}$ ($\sigma_{l}^{-}$) creates
(annihilates) a particle at site $l$.
Thus the terms $\sigma_{l}^{-}\sigma_{l+1}^{+}$ and
$\sigma_{l}^{+}\sigma_{l+1}^{-}$ generate particle hopping to
the right and left respectively --- they derive from the gain
terms of the original Master equation (i.e. the terms due to particle
hopping which increase the probability of finding the system in a
particular configuration). The terms involving projection
operators arise due to the loss terms in the Master equation (i.e. the
terms which contribute to the probability that the system is in a
particular configuration provided that no particle performs a hop). 
Therefore probability and particle number (which is related to the
$z$-component of the spin) are both conserved. This model also possesses a
particle-hole symmetry under interchange of $p \leftrightarrow q$. The
aim is to maintain this duality at all stages of scaling. 

We begin by dividing the lattice into blocks of size $b=2$. Then we
split $H$ into an intrablock Hamiltonian $H_{\nu}$ given by
\begin{equation}
H_{\nu} = p\:(P_{\nu,1}^{+}P_{\nu,2}^{-}-\sigma_{\nu,1}^{-}\sigma_{\nu,2}^{+})
+q\:(P_{\nu,1}^{-}P_{\nu,2}^{+}-\sigma_{\nu,1}^{+}\sigma_{\nu,2}^{-}) , 
\end{equation}
where the suffix $\nu,i$ indicates that the operator acts at site $i$
in block $\nu$, and an interblock Hamiltonian $H_{\nu,\nu +1}$ given by
\begin{equation}
H_{\nu,\nu +1} = p\:(P^{+}_{\nu,2}P_{\nu+1,1}^{-}-\sigma_{\nu,2}^{-}\sigma_{\nu+1,1}^{+})
+q\:(P_{\nu,2}^{-}P_{\nu+1,1}^{+}-\sigma_{\nu,2}^{+}\sigma_{\nu+1,1}^{-}) ,
\end{equation}
such that both $H_{\nu}$ and $H_{\nu,\nu +1}$ possess the duality
under interchange of $p \leftrightarrow q$ when $+ \leftrightarrow -$.

The next step is to find the lowest lying eigenstates of $H_{\nu}$ and use
them to form the renormalised basis of spin states. The ground state
of $H_{\nu}$ is three-fold degenerate --- since the dynamics conserve
particle number $H_\nu$ 
decomposes into $b+1$ disconnected sectors. This particle conservation
can be respected in the renormalised Hamiltonian if it is present
in the renormalised basis of states, therefore the ground eigenstates
of $H_\nu$ are organised
according to their eigenvalue of the block spin operator
$\frac{1}{b}\sum_{i=1}^b \sigma_{\nu,i}^z$. This leads us to define
block spin states  
\begin{eqnarray}
&| +1_{\nu} \rangle = |+_{\nu,1},+_{\nu,2}\rangle \;, \\
&| 0_{\nu} \rangle = \frac{1}{p+q} [ q \vert
+_{\nu,1},-_{\nu,2}\rangle + p \vert -_{\nu,1},+_{\nu,2}\rangle ] \;,\\
&| -1_{\nu} \rangle = |-_{\nu,1},-_{\nu,2}\rangle \;,
\end{eqnarray}
for each block $\nu$. The corresponding left eigenstates of $H_\nu$
are assigned to the left block spin states
\begin{eqnarray}
&\langle +1_{\nu} \vert = \langle +_{\nu,1},+_{\nu,2}\vert \;, \\
&\langle 0_{\nu} \vert = \langle +_{\nu,1},-_{\nu,2}
\vert + \langle -_{\nu,1},+_{\nu,2}\vert  \;,\\
&\langle -1_{\nu} \vert = \langle -_{\nu,1},-_{\nu,2}\vert \;.
\end{eqnarray}
The left ground eigenstates of a quantum Hamiltonian describing a
nonequilibrium process are always given by such sums over vectors
(where the coefficient of each vector is equal to one) due to
conservation of probability.   
By defining the block states in this way, we aim to maintain particle
conservation (which we cannot maintain by arranging these states in two
linear superpositions forming a spin-1/2 basis) and the particle-hole
duality of the model. Thus the renormalised Hamiltonian will describe
a spin-1 quantum chain. 

We are now able to calculate the matrix elements of $H'_\nu$ and
$H'_{\nu,\nu+1}$. Firstly, we note that because our basis states are
degenerate eigenstates of $H_\nu$, the contribution due to $H'_\nu$ is 
a constant and moreover, because the groundstate eigenvalue is zero
(which is always the case for the ground state eigenvalue of quantum
systems representing nonequilibrium models) this constant is zero. The
task then is to evaluate the matrix elements of
$H'_{\nu,\nu+1}$, as prescribed by equation (\ref{V}). For example,
the term in $H_{\nu,\nu+1}$ given by $\sigma_{\nu,2}^-
\sigma_{\nu+1,1}^+$ contributes the matrix elements
\begin{equation}
\langle \sigma_\nu , \sigma_{\nu+1} \vert H'_{\nu,\nu+1} \vert
 \sigma'_\nu , \sigma'_{\nu+1} \rangle = \langle
 \sigma_\nu \vert \sigma_{\nu,2}^- \vert \sigma'_\nu \rangle \langle
 \sigma_{\nu+1} \vert \sigma_{\nu+1,1}^+ \vert \sigma'_{\nu+1} \rangle \;.
\end{equation}
Thus the operator $\sigma_{\nu,2}^-$ is replaced by a renormalised
operator given by  
\begin{displaymath}
\sigma_{\nu,2}^- = \left( \begin{array}{ccc}
\langle +1_\nu \vert \sigma_{\nu,2}^- \vert +1_\nu \rangle &
\langle +1_\nu \vert \sigma_{\nu,2}^- \vert 0_\nu \rangle &
\langle +1_\nu \vert \sigma_{\nu,2}^- \vert -1_\nu \rangle \\
\langle 0_\nu \vert \sigma_{\nu,2}^- \vert +1_\nu \rangle &
\langle 0_\nu \vert \sigma_{\nu,2}^- \vert 0_\nu \rangle &
\langle 0_\nu \vert \sigma_{\nu,2}^- \vert -1_\nu \rangle \\
\langle -1_\nu \vert \sigma_{\nu,2}^- \vert +1_\nu \rangle &
\langle -1_\nu \vert \sigma_{\nu,2}^- \vert 0_\nu \rangle &
\langle -1_\nu \vert \sigma_{\nu,2}^- \vert -1_\nu \rangle \end{array}
\right)_\nu \;,
\end{displaymath}
where the $\nu$ suffix on the matrix represents an operator acting on
the truncated basis of states in block $\nu$. Evaluating such matrix
elements for all the operators appearing in $H_{\nu,\nu+1}$ yields 
\begin{eqnarray}
P_{\nu,2}^+ = P_{\nu}^{+\!1} + \textstyle\frac{p}{p+q}P_{\nu}^0 , \quad
P_{\nu+1,1}^+ = P_{\nu+1}^{+\!1} + \textstyle\frac{q}{p+q}P_{\nu+1}^0 ,
\nonumber \\ 
P_{\nu,2}^- = \textstyle\frac{q}{p+q}P_{\nu}^0 + P_{\nu}^{-\!1} , \quad
P_{\nu+1,1}^- = \textstyle\frac{p}{p+q}P_{\nu+1}^0 + P_{\nu+1}^{-\!1},
\nonumber \\ 
\sigma_{\nu,2}^+ = \textstyle\frac{q}{p+q}\sigma_{\nu}^{0,+\!1} +
\sigma_{\nu}^{-\!1,0} , \quad 
\sigma_{\nu+1,1}^+ = \textstyle\frac{p}{p+q}\sigma_{\nu+1}^{0,+\!1} +
\sigma_{\nu+1}^{-\!1,0}, \nonumber \\ 
\sigma_{\nu,2}^- = \sigma_{\nu}^{+\!1,0} +
\textstyle\frac{p}{p+q}\sigma_{\nu}^{0,-\!1} , \quad  
\sigma_{\nu+1,1}^- = \sigma_{\nu+1}^{+\!1,0} +
\textstyle\frac{q}{p+q}\sigma_{\nu+1}^{0,-\!1} , \nonumber 
\end{eqnarray} 
where $P_{\nu}^i$ projects into the block spin state $i$ and
$\sigma_{\nu}^{i,j}$ raises or lowers the spin from block spin state
$i$ to block spin
state $j$. These expressions are substituted into $H_{\nu,\nu+1}$
leading to a  renormalised Hamiltonian given by $H' = \sum_\nu
H_{\nu,\nu+1}$.

Thus the renormalised Hamiltonian describes a three-state stochastic
process (probability is still conserved) where the $z$-component of
the spin is still conserved. In order to obtain scaling
equations for the rates $p$ and $q$ this Hamiltonian has to be
projected onto a basis of spin-1/2 states. To do this, we note that
raising (lowering) operators in the spin-1/2 system are
written only in terms of operators which raise (lower) the spin in the
spin-1 system. Further, each of the dynamical processes, generated by
the raising and lowering operators in the spin-1 system, can be
identified with a corresponding term involving projection operators due to
no transition. Hence we rewrite
\begin{eqnarray}
P_{\nu,2}^+ = (1 + \textstyle\frac{p}{p+q})P_{\nu}^+ , \quad
P_{\nu+1,1}^+ = (1 + \textstyle\frac{q}{p+q})P_{\nu+1}^+ ,
\nonumber \\ 
P_{\nu,2}^- = (1+\textstyle\frac{q}{p+q})P_{\nu}^-  , \quad
P_{\nu+1,1}^- = (1+\textstyle\frac{p}{p+q})P_{\nu+1}^- , \nonumber \\ 
\sigma_{\nu,2}^+ = (1+\textstyle\frac{q}{p+q})\sigma_{\nu}^+ , \quad 
\sigma_{\nu+1,1}^+ = (1+\textstyle\frac{p}{p+q})\sigma_{\nu+1}^+ ,
\nonumber \\  
\sigma_{\nu,2}^- = (1+\textstyle\frac{p}{p+q})\sigma_{\nu}^- , \quad  
\sigma_{\nu+1,1}^- = (1+\textstyle\frac{q}{p+q})\sigma_{\nu+1}^- . \nonumber 
\end{eqnarray} 
whence the renormalised Hamiltonian assumes the same form as the
original: 
\begin{equation} 
H' = \sum_{\nu} \left[
p'\:(P_{\nu}^{+}P_{\nu+1}^{-}-\sigma_{\nu}^{-}\sigma_{\nu+1}^{+}) 
+q'\:(P_{\nu}^{-}P_{\nu+1}^{+}-\sigma_{\nu}^{+}\sigma_{\nu+1}^{-}) \right] \;, 
\end{equation}
where the rescaled rates $p'$ and $q'$ are given by
\begin{equation}
p' = p(1 + \textstyle\frac{p}{p+q})^2 , \quad q'= q(1 +
\textstyle\frac{q}{p+q})^2 \;.
\end{equation}

In order to exploit this renormalisation, we consider the ratio
$\gamma=p/q$. Stable fixed points for $\gamma$ are
found at $\gamma^* = 0$ and $\infty$, and these are separated by an
unstable fixed point at $\gamma^* = 1$. Hence symmetric diffusion is
unstable with respect to bias. At the symmetric fixed
point the Hamiltonian (\ref{quant_ASEP}) describes a spin-1/2
Heisenberg chain whose dynamics are governed by a dynamic exponent
$z=2$. This behaviour therefore is unstable and in the presence of
any bias the dynamics are described by a new exponent. A Bethe ansatz
calculation for $q=0$ shows that this exponent is $z=3/2$ \cite{gwa92}.
That we find the exact value for the
unstable fixed point is a consequence of our preservation of duality
at all stages in the blocking.


\subsection{Scaling for the Pair Evaporation and Deposition Model}
\label{sect:quant_PED}

Another model possessing a duality is a process whereby pairs of
particles evaporate from adjacent lattice sites with a rate
$\epsilon$ or are deposed onto adjacent vacancies with a rate
$\delta$. The quantum Hamiltonian representing these processes takes
the form
\begin{equation} \label{PED}
H = \sum_{l} \left[  
\delta \left( P_{l}^{-}P_{l+1}^{-}-\sigma_{l}^{+}\sigma_{l+1}^{+} \right)
+\epsilon \left( P_{l}^{+}P_{l+1}^{+}-\sigma_{l}^{-}\sigma_{l+1}^{-} \right) 
 \right] \;.
\end{equation}
The duality in this model is again a particle-hole symmetry under the
interchange of $\delta \leftrightarrow \epsilon$. There is also a
conservation law similar to the particle conservation in the ASEP: if
we label the sublattice of odd (even) sites by A (B), then the density
of particles on sublattice A (B) is $\rho_A$ ($\rho_B$). The conserved
quantity is the difference in the sublattice densities, $\rho_A-\rho_B$.

The duality of this Hamiltonian is preserved if we divide it into an
intrablock part given by
\begin{equation}
 H_{\nu} = 
\delta \:(P_{\nu,1}^{-}P_{\nu,2}^{-}-\sigma_{\nu,1}^{+}\sigma_{\nu,2}^{+})
+\epsilon
\:(P_{\nu,1}^{+}P_{\nu,2}^{+}-\sigma_{\nu,1}^{-}\sigma_{\nu,2}^{-})
\;, 
\end{equation}
and an interblock part given by
\begin{equation}
  H_{\nu,\nu+1} = 
\delta \:(P_{\nu,2}^{-}P_{\nu+1,1}^{-}-\sigma_{\nu,2}^{+}\sigma_{\nu+1,1}^{+})
+\epsilon
\:(P_{\nu,2}^{+}P_{\nu+1,1}^{+}-\sigma_{\nu,2}^{-}\sigma_{\nu+1,1}^{-})
\;. 
\end{equation}
Again, the ground state of $H_{\nu}$ is three-fold degenerate with
eigenvectors given by
\begin{eqnarray}
| 1_{\nu} \rangle &=& |-_{\nu,1},+_{\nu,2}\rangle \;, \\
| 2_{\nu} \rangle &=& \frac{1}{\delta+\epsilon} [
 \delta \vert +_{\nu,1},+_{\nu,2}\rangle +
 \epsilon |-_{\nu,1},-_{\nu,2}\rangle ] \;,\\ 
| 3_{\nu} \rangle &=& |+_{\nu,1},-_{\nu,2}\rangle \;.
\end{eqnarray}
We can maintain the conservation of the sublattice densities with
respect to both the intrablock \emph{and} the interblock Hamiltonians
if we assign spin-1 block states in the following way:
\begin{eqnarray}
&|+1_{\nu}\rangle = |1_{\nu}\rangle \;, \quad &|+1_{\nu+1}\rangle =
 |3_{\nu+1}\rangle \;, \nonumber \\
&|0_{\nu}\rangle = |2_{\nu}\rangle \;, \quad &|0_{\nu+1}\rangle =
 |2_{\nu+1}\rangle \;,\\
&|-1_{\nu}\rangle = |3_{\nu}\rangle \;, \quad &|-1_{\nu+1}\rangle =
 |1_{\nu+1}\rangle \;. \nonumber
\end{eqnarray}
Using these states to calculate $H'_{\nu,\nu+1}$
($H'_\nu$ is zero as before), and then forcing the resulting spin-1
Hamiltonian back into a spin-1/2 basis in the same way as was done in the
previous section, one obtains a renormalised Hamiltonian of the
same form as (\ref{PED}) with rescaled rates $\delta'$ and
$\epsilon'$ given by  
\begin{equation}
\delta' = \delta(1 +
\textstyle\frac{\epsilon}{\delta+\epsilon})^2 \;, \quad
\epsilon' = \epsilon(1 +
\textstyle\frac{\delta}{\delta+\epsilon})^2 \;.
\end{equation}
Now, in terms of the ratio $\gamma \equiv \delta / \epsilon$,
we again find three fixed points at $\gamma^* = 0, 1$ and $\infty$ but
now the symmetric fixed point $\gamma^* = 1$ is stable. Therefore
we find no dynamic transition in this model --- the dynamics are
independent of bias. Again, the Hamiltonian
(\ref{PED}) of the symmetric problem is given by that of the
spin-1/2 Heisenberg chain. Thus the dynamic exponent is $z=2$
in the biased and unbiased cases. This result is consistent
with a mean field result which predicts that the diffusion constant is
independent of bias, and it is supported by numerical simulation
\cite{grynberg94}. 


\subsection{Scaling for the Pair Evaporation and Deposition Process
with Diffusion in One Dimension} \label{subsect:1dpca}

In this section, we apply the renormalisation group transformation to
a model incorporating both the dynamics of the ASEP and of the pair
evaporation and deposition process.  
These dynamics, on a chain with periodic boundary
conditions, are represented by a quantum Hamiltonian given by
\begin{eqnarray} \label{EDDH}
  H = \sum_{l} &\left[  
\delta \left( P_{l}^{-}P_{l+1}^{-}-\sigma_{l}^{+}\sigma_{l+1}^{+} \right)
+\epsilon \left( P_{l}^{+}P_{l+1}^{+}-\sigma_{l}^{-}\sigma_{l+1}^{-} \right) 
 \right. \nonumber \\
&\left.\,+\,p\left( P_{l}^{+}P_{l+1}^{-}-\sigma_{l}^{-}\sigma_{l+1}^{+}\right)
+q\left(P_{l}^{-}P_{l+1}^{+}-\sigma_{l}^{+}\sigma_{l+1}^{-}\right) \right] \;. 
\end{eqnarray}
Exact results have
been obtained when $\delta = \epsilon$ and $p=q$ in which case the
model is equivalent to a spin-1/2 XXZ ferromagnet, and
also for $\delta + \epsilon = p+q$ which is the condition that the
evolution operator can be written as a free-fermion Hamiltonian 
\cite{grynberg94,grynberg95}. 

A blocking, with a dilation factor $b=2$, is implemented by dividing
$H$ into an intrablock part $H_{\nu}$ given by 
\begin{eqnarray}
  H_{\nu} = 
&\delta \:(P_{\nu,1}^{-}P_{\nu,2}^{-}-\sigma_{\nu,1}^{+}\sigma_{\nu,2}^{+})
+\epsilon\:(P_{\nu,1}^{+}P_{\nu,2}^{+}-\sigma_{\nu,1}^{-}\sigma_{\nu,2}^{-})
 \nonumber \\
&+p\:(P_{\nu,1}^{+}P_{\nu,2}^{-}-\sigma_{\nu,1}^{-}\sigma_{\nu,2}^{+})
+q\:(P_{\nu,1}^{-}P_{\nu,2}^{+}-\sigma_{\nu,1}^{+}\sigma_{\nu,2}^{-}) \;, 
\end{eqnarray}
and an interblock part given by
\begin{eqnarray}
  H_{\nu,\nu+1} = 
&\delta\:(P_{\nu,2}^{-}P_{\nu+1,1}^{-}-\sigma_{\nu,2}^{+}\sigma_{\nu+1,1}^{+})
+\epsilon\:(P_{\nu,2}^{+}P_{\nu+1,1}^{+}-\sigma_{\nu,2}^{-}\sigma_{\nu+1,1}^{-}) \nonumber \\ 
&+p\:(P^{+}_{\nu,2}P_{\nu+1,1}^{-}-\sigma_{\nu,2}^{-}\sigma_{\nu+1,1}^{+})
+q\:(P_{\nu,2}^{-}P_{\nu+1,1}^{+}-\sigma_{\nu,2}^{+}\sigma_{\nu+1,1}^{-})\;.  
\end{eqnarray}
The blocking proceeds as in the previous two sections. 
The ground state of $H_{\nu}$ is two-fold degenerate with eigenvectors denoted
\begin{eqnarray}
  |1_{\nu}\rangle = & \frac{1}{\delta+\epsilon}
\left[\delta|+_{\nu,1}+_{\nu,2}\rangle+\epsilon|-_{\nu,2}-_{\nu,2}\rangle\right]\;, \\ 
  |2_{\nu}\rangle = & \frac{1}{p+q}
\left[q|+_{\nu,1}-_{\nu,2}\rangle+p|-_{\nu,1}+_{\nu,2}\rangle\right] \;.
\end{eqnarray}
Effective spin states for the block $\nu$ are identified by taking the
block up-spin $| +_{\nu} \rangle = |1_{\nu}\rangle$ and the block
down-spin $|-_{\nu}\rangle = |2_{\nu}\rangle$ for all $\nu$. Thus the
block spin states
observe two symmetries of the model:
the particle-hole symmetry under interchange of rates
$\delta \leftrightarrow \epsilon$ and $p \leftrightarrow q$, and a
symmetry whereby the spins on the even sublattice (say) of sites
are flipped and the pair evaporation and deposition processes are
transformed into the hopping processes and vice versa. This assignment
of block spin states yields a renormalised Hamiltonian $H'$ of the same form as
(\ref{EDDH}), but with rescaled rates $\delta'$,$\epsilon'$,$p'$ and
$q'$ given by
\begin{eqnarray}
\delta' = \frac{p^3+q^3+pq(\delta+\epsilon)}{(p+q)^2} , \\
\epsilon' =
\frac{\delta\epsilon(\delta+\epsilon+p+q)}{(\delta+\epsilon)^2}, 
\\
p' = \frac{\delta\epsilon (p+q)+\delta p^2+\epsilon
q^2}{(\delta+\epsilon)(p+q)} , \\
q' = \frac{\delta\epsilon (p+q)+\epsilon p^2+\delta
q^2}{(\delta+\epsilon)(p+q)}. 
\end{eqnarray}
A flow diagram is obtained where the symmetric fixed point (i.e. where
all rates are equal) is fully stable. This
suggests that the dynamic transition in the ASEP is removed when the
pair evaporation and deposition processes are included --- the dynamics
in this model are described by the exponent $z=2$ for all choices of
$\delta$, $\epsilon$, $p$ and $q$ (provided $\delta$ and
$\epsilon$ are not both equal to zero). Also, all the fixed points
satisfy the free-fermion condition $\delta + \epsilon = p+q$ for
which exact results are available
\cite{grynberg95,grynberg94}. Moreover, this condition is stable
(after iterating the transformation an infinite number of times the
rescaled rates always satisfy the free-fermion condition, regardless
of the original choice of rates), therefore we do not expect to
observe any new macroscopic behaviour in the regions of
parameter-space to which the exact solution does not apply.


\section{Trotter Decomposition for the Asymmetric Simple Exclusion
Process}

One difficulty that arises when constructing
renormalisation group transformations for quantum systems is
associated with the non-commutation of operators appearing in the
quantum Hamiltonian. In this section, we remove this problem by
exploiting the Suzuki-Trotter decomposition
\cite{trotter59,suzuki76,suzuki77,suzuki77a} to rewrite the quantum
Hamiltonian representing the ASEP as a classical Hamiltonian for Ising
spin variables in two dimensions. As a byproduct, direct contact is
made between stochastic nonequilibrium models in one dimension and
vertex models in two dimensions \cite{baxter82}.

The general scheme for the mapping is to split the Hamiltonian into a set
of operators $\{ H_i \}$ such that
\begin{equation}
H = \sum_{i=0}^{j}\,H_i,
\end{equation}
where each term in $H_i$ commutes with every other (or at least, where
any non-commutation can be easily dealt with), but  
$[ H_l , H_m] \neq 0$ for $l \neq m$. 
The exponential of a sum of operators is then expanded as a product of
exponentials by dealing with the non-commutation through the Trotter
formula \cite{trotter59} 
\begin{equation} \label{trotter_form}
e^{\sum_{i=0}^{j} H_i} = \lim_{n\rightarrow \infty} 
  \left[ e^{\frac{H_0}{n}} \ldots e^{\frac{H_j}{n}} \right]^n .
\end{equation}
 
For the ASEP, we begin with the quantum Hamiltonian
(\ref{quant_ASEP}), and divide it up in the following way:
\begin{eqnarray}
H_0 &= \sum_{l \ {\rm odd}} 
\left[ p\:P_{l}^{+}P_{l+1}^{-}+q\:P_{l}^{-}P_{l+1}^{+} \right],
\label{H0} \\
H_1 &= -\sum_{l \ {\rm odd}} \left[ p\:\sigma_{l}^{-}\sigma_{l+1}^{+}+q\:\sigma_{l}^{+}\sigma_{l+1}^{-} \right], \\
H_2 &= \sum_{l \ {\rm even}} 
\left[ p\:P_{l}^{+}P_{l+1}^{-}+q\:P_{l}^{-}P_{l+1}^{+} \right], \\
H_3 &= -\sum_{l \ {\rm even}} \left
[ p\:\sigma_{l}^{-}\sigma_{l+1}^{+}+q\:\sigma_{l}^{+}\sigma_{l+1}^{-}
\right]. \label{H3}
\end{eqnarray}
Ultimately, this choice must be made to reflect the update mechanism ---
here we consider a parallel sublattice update. According to equation
(\ref{trotter_form}) the partition function for the quantum system is
written    
\begin{equation}
Z = \lim_{n \rightarrow \infty} Z_{(n)},
\end{equation}
where we have defined $Z_{(n)}$ by
\begin{equation} \label{Zn}
 Z_{(n)} \equiv {\rm Tr} 
\left[ e^{-\frac{\beta H_0}{n}} e^{-\frac{\beta H_1}{n}}
e^{-\frac{\beta H_2}{n}} e^{-\frac{\beta H_3}{n}} \right]^n .
\end{equation}
The next step is to insert complete sets of basis states into
(\ref{Zn}). We choose the basis $\vert s \rangle = \vert
\sigma_1,\sigma_2,\ldots , \sigma_L \rangle$, where $\sigma_i$ is the
eigenvalue of the Pauli matrix $\sigma_i^z$; a basis which
diagonalises $H_0$ and $H_2$. Hence $Z_{(n)}$ is expressed as follows
\begin{eqnarray}
Z_{(n)} \!\!\!&=&\!\!\! \sum_{\{ s_i \} } 
\langle s_{1} \vert e^{-\frac{\beta H_0}{n}} e^{-\frac{\beta H_1}{n}} \vert s_2 \rangle  
\langle s_{2} \vert e^{-\frac{\beta H_2}{n}} e^{-\frac{\beta H_3}{n}}
\vert s_3 \rangle \ldots 
\langle s_{2n} \vert e^{-\frac{\beta H_2}{n}} e^{-\frac{\beta H_3}{n}}
\vert s_1 \rangle \nonumber \\
\!\!\!&=&\!\!\! \sum_{\{ s_i \} } 
\left[ \prod_{\tau \ {\rm odd}} \langle s_{\tau} \vert e^{-\frac{\beta H_0}{n}} e^{-\frac{\beta H_1}{n}} \vert s_{\tau +1} \rangle \right]\!\!
\left[ \prod_{\tau \ {\rm even}} \langle s_{\tau} \vert
e^{-\frac{\beta H_2}{n}} e^{-\frac{\beta H_3}{n}} \vert s_{\tau +1}
\rangle \right] . \label{insertion}
\end{eqnarray}
This insertion of basis states also reflects the update mechanism of
the original particle process. Now we interpret the $\tau$ label as a
label for a new spatial axis, which we will refer to as the Trotter
axis (see Figure \ref{fig:ising_lattice}). This axis has its origin in
the time axis of the quantum representation of the Master
equation. Usually, the Trotter axis is an imaginary time
dimension, but since our quantum problem is governed by a
Schr\"odinger equation in imaginary time, here the Trotter axis
represents a real time evolution axis. 

From now, for simplicity, we shall consider the fully ASEP: $q=0$. Our
task now is to evaluate the matrix elements appearing in
(\ref{insertion}). 
Since we chose our basis states to diagonalise $H_0$, the contribution
from its exponential factor is trivially evaluated. 
The contribution from the off-diagonal part, $H_1$ proceeds as follows:
\begin{equation}
\prod_{\tau \ {\rm odd}} \langle s_{\tau} \vert {\rm exp} \left[
\sum_{l \ {\rm odd}} 
\left( \frac{\beta p}{n} \sigma_{l}^{-}\sigma_{l+1}^{+} \right)
\right] \vert s_{\tau+1} \rangle = \lim_{\Delta \rightarrow \infty} {\rm
exp} \left(-\sum_{l,\tau \ {\rm odd}} h_{l,\tau} \right) ,
\end{equation}
where $h_{l,\tau}$ is given by
\begin{equation}
h_{l,\tau} = \Delta \Phi_{l,\tau} -\ln\left( \frac{\beta p}{n}
\right) P_{l,\tau}^{+} P_{l,\tau +1}^{-} P_{l+1,\tau}^{-} P_{l+1,\tau
+1}^{+} ,
\end{equation}
with $P_{l,\tau}^{\pm} = \frac{1}{2} (1 \pm \sigma_{l,\tau})$ and
$\sigma_{l,\tau}$ is an Ising spin variable associated with the site
$(l,\tau)$, and where
\begin{eqnarray} \label{Phi_def}
\Phi_{l,\tau} = & 
P_{l,\tau}^{+} P_{l,\tau +1}^{+} P_{l+1,\tau}^{+} P_{l+1,\tau +1}^{-}
+P_{l,\tau}^{+} P_{l,\tau +1}^{+} P_{l+1,\tau}^{-} P_{l+1,\tau +1}^{+} 
\nonumber \\
&+P_{l,\tau}^{+} P_{l,\tau +1}^{-} P_{l+1,\tau}^{+} P_{l+1,\tau +1}^{+} 
+P_{l,\tau}^{-} P_{l,\tau +1}^{+} P_{l+1,\tau}^{+} P_{l+1,\tau +1}^{+} 
\nonumber \\
&+P_{l,\tau}^{+} P_{l,\tau +1}^{-} P_{l+1,\tau}^{-} P_{l+1,\tau +1}^{-}
+P_{l,\tau}^{-} P_{l,\tau +1}^{+} P_{l+1,\tau}^{-} P_{l+1,\tau +1}^{-}
\nonumber \\
&+P_{l,\tau}^{-} P_{l,\tau +1}^{-} P_{l+1,\tau}^{+} P_{l+1,\tau +1}^{-}
+P_{l,\tau}^{-} P_{l,\tau +1}^{-} P_{l+1,\tau}^{-} P_{l+1,\tau +1}^{+}
\nonumber \\
&+P_{l,\tau}^{+} P_{l,\tau +1}^{-} P_{l+1,\tau}^{+} P_{l+1,\tau +1}^{-} 
+P_{l,\tau}^{-} P_{l,\tau +1}^{+} P_{l+1,\tau}^{-} P_{l+1,\tau +1}^{+} 
\nonumber \\ 
&+P_{l,\tau}^{-} P_{l,\tau +1}^{+} P_{l+1,\tau}^{+} P_{l+1,\tau
+1}^{-}. 
\end{eqnarray}
The parameter $\Delta$ has been introduced in order to project away
unwanted configurations, i.e.\ those which do not represent processes
allowed under the original stochastic dynamics; for instance, the
final term in equation (\ref{Phi_def}) corresponds to a particle
hopping to the left, which is forbidden for $q=0$.

To obtain the full effective Hamiltonian for the two-dimensional classical
system, $Z_{(n)} = \lim_{\Delta \rightarrow \infty} {\rm Tr}
e^{-H^{(\mathit{eff})}}$, we must include the contributions from the
the Hamiltonians $H_0$, $H_2$ and $H_3$. Then the effective Hamiltonian can
be written  
\begin{equation} \label{class_ASEP1}
H^{(\mathit{eff})} = \sum_{l,\tau \ {\rm odd}}
h_{l,\tau}^{(\mathit{eff})} + \sum_{l,\tau \ {\rm even}}
h_{l,\tau}^{(\mathit{eff})} , 
\end{equation}
with
\begin{eqnarray} \label{class_ASEP2}
h_{l,\tau}^{(\mathit{eff})} =& \Delta \Phi_{l,\tau}
-\ln\left( \frac{\beta p}{n} \right) 
P_{l,\tau}^{+} P_{l,\tau +1}^{-} P_{l+1,\tau}^{-} P_{l+1,\tau +1}^{+} 
\nonumber \\
&- \ln\left( 1- \frac{\beta p}{n} \right) P_{l,\tau}^{+} P_{l,\tau
+1}^{+} P_{l+1,\tau}^{-} P_{l+1,\tau +1}^{-} ,
\end{eqnarray}
where the final term here corresponds to the `stay-put' probability
that a particle, with a vacancy to its right, does not perform a hop. 

The quantum Hamiltonian (\ref{quant_ASEP}) has now been rewritten as a
classical Hamiltonian (\ref{class_ASEP1}) for
an Ising system in two dimensions. The nonequilibrium steady state of
the original model is characterised by the zero temperature behaviour
of the quantum model. This limit is recovered in the Ising system by
taking the extent of the Trotter axis to be infinite. The
temperature in the quantum system is not the same as the temperature
in the classical Ising system. Instead, the inverse temperature of the quantum
model translates into the extent of the classical system in the
Trotter direction. Thus the
critical behaviour in the quantum ground state is expressed through
the critical behaviour of the finite temperature Ising system in
equilibrium. Zero temperature transitions in the quantum model, which
occur as a function of the couplings, are caused  in the infinite
Ising system by varying the temperature.


\subsection{`Plaquette' Structure of Ising Hamiltonian}

\begin{figure}
\begin{center}
\unitlength=1mm
\linethickness{0.1pt}
\begin{picture}(57,12)

\put(1,5){\line(1,0){8}}
\put(1,11){\line(1,0){8}}
\put(2,4){\line(0,1){8}}
\put(8,4){\line(0,1){8}}
\put(1.17,4.3){$\uparrow$}
\put(1.17,10.3){$\uparrow$}
\put(7.17,4.3){$\uparrow$}
\put(7.17,10.3){$\uparrow$}
\put(4,0){$\omega_1$}

\put(13,5){\line(1,0){8}}
\put(13,11){\line(1,0){8}}
\put(14,4){\line(0,1){8}}
\put(20,4){\line(0,1){8}}
\put(13.17,4.3){$\downarrow$}
\put(13.17,10.3){$\downarrow$}
\put(19.17,4.3){$\downarrow$}
\put(19.17,10.3){$\downarrow$}
\put(16,0){$\omega_2$}

\put(25,5){\line(1,0){8}}
\put(25,11){\line(1,0){8}}
\put(26,4){\line(0,1){8}}
\put(32,4){\line(0,1){8}}
\put(25.17,4.3){$\uparrow$}
\put(25.17,10.3){$\uparrow$}
\put(31.17,4.3){$\downarrow$}
\put(31.17,10.3){$\downarrow$}
\put(28,0){$\omega_3$}

\put(37,5){\line(1,0){8}}
\put(37,11){\line(1,0){8}}
\put(38,4){\line(0,1){8}}
\put(44,4){\line(0,1){8}}
\put(37.17,4.3){$\downarrow$}
\put(37.17,10.3){$\downarrow$}
\put(43.17,4.3){$\uparrow$}
\put(43.17,10.3){$\uparrow$}
\put(40,0){$\omega_4$}

\put(49,5){\line(1,0){8}}
\put(49,11){\line(1,0){8}}
\put(50,4){\line(0,1){8}}
\put(56,4){\line(0,1){8}}
\put(49.17,4.3){$\uparrow$}
\put(49.17,10.3){$\downarrow$}
\put(55.17,4.3){$\downarrow$}
\put(55.17,10.3){$\uparrow$}
\put(52,0){$\omega_5$}

\end{picture}
\caption{The allowed plaquette configurations with their weights} 
\label{fig:vertices}
\end{center}
\end{figure}
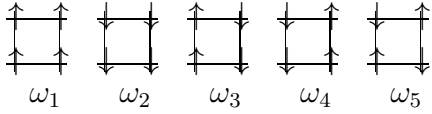

The Ising Hamiltonian $H^{(\mathit{eff})}$ representing the ASEP
contains a parameter $\Delta$ which is 
infinite. This constraint can be incorporated in a compact fashion
into a vertex model description. In the representation we have
chosen, $H^{(\mathit{eff})}$ contains only four-spin
interactions. These are shown in Figure \ref{fig:vertices} as the 
five possible arrangements of spins around a plaquette.

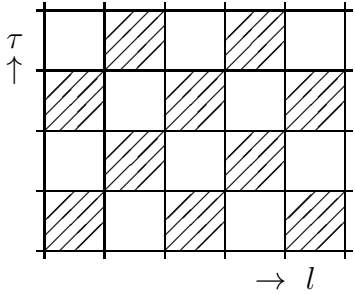
\begin{figure}
\begin{center}
\unitlength=1mm
\linethickness{0.4pt}
\begin{picture}(46,38)

\put(4,5){\line(1,0){42}}
\put(4,13){\line(1,0){42}}
\put(4,21){\line(1,0){42}}
\put(4,29){\line(1,0){42}}
\put(4,37){\line(1,0){42}}

\put(5,4){\line(0,1){34}}
\put(13,4){\line(0,1){34}}
\put(21,4){\line(0,1){34}}
\put(29,4){\line(0,1){34}}
\put(37,4){\line(0,1){34}}
\put(45,4){\line(0,1){34}}

\put(5,5){\line(1,1){8}}
\put(5,7){\line(1,1){6}}
\put(5,9){\line(1,1){4}}
\put(7,5){\line(1,1){6}}
\put(9,5){\line(1,1){4}}
\put(5,21){\line(1,1){8}}
\put(5,23){\line(1,1){6}}
\put(5,25){\line(1,1){4}}
\put(7,21){\line(1,1){6}}
\put(9,21){\line(1,1){4}}
\put(21,5){\line(1,1){8}}
\put(21,7){\line(1,1){6}}
\put(21,9){\line(1,1){4}}
\put(23,5){\line(1,1){6}}
\put(25,5){\line(1,1){4}}
\put(21,21){\line(1,1){8}}
\put(21,23){\line(1,1){6}}
\put(21,25){\line(1,1){4}}
\put(23,21){\line(1,1){6}}
\put(25,21){\line(1,1){4}}
\put(37,5){\line(1,1){8}}
\put(37,7){\line(1,1){6}}
\put(37,9){\line(1,1){4}}
\put(39,5){\line(1,1){6}}
\put(41,5){\line(1,1){4}}
\put(37,21){\line(1,1){8}}
\put(37,23){\line(1,1){6}}
\put(37,25){\line(1,1){4}}
\put(39,21){\line(1,1){6}}
\put(41,21){\line(1,1){4}}
\put(13,13){\line(1,1){8}}
\put(13,15){\line(1,1){6}}
\put(13,17){\line(1,1){4}}
\put(15,13){\line(1,1){6}}
\put(17,13){\line(1,1){4}}
\put(13,29){\line(1,1){8}}
\put(13,31){\line(1,1){6}}
\put(13,33){\line(1,1){4}}
\put(15,29){\line(1,1){6}}
\put(17,29){\line(1,1){4}}
\put(29,13){\line(1,1){8}}
\put(29,15){\line(1,1){6}}
\put(29,17){\line(1,1){4}}
\put(31,13){\line(1,1){6}}
\put(33,13){\line(1,1){4}}
\put(29,29){\line(1,1){8}}
\put(29,31){\line(1,1){6}}
\put(29,33){\line(1,1){4}}
\put(31,29){\line(1,1){6}}
\put(33,29){\line(1,1){4}}

\put(33,0){$\rightarrow \ l$}
\put(0,28){$\uparrow$}
\put(0,32){$\tau$}

\end{picture}
\caption{Lattice on which the 2$d$ classical Ising model is
defined. The original spatial axis is labelled by $l$ and $\tau$ labels
the Trotter axis. The configurations around the shaded squares are
determined by the allowed plaquette configurations shown in Figure
\ref{fig:vertices}.}  \label{fig:ising_lattice}
\end{center}
\end{figure}

The weights $\omega_1$ to $\omega_5$ of the allowed vertices are
\begin{equation} 
\omega_1 = \omega_2 = \omega_4 = 1 , \nonumber \\
\omega_3 = 1-\frac{\beta p}{n} ,\\
\omega_5 = \frac{\beta p}{n}. \nonumber
\end{equation}
All other vertices have zero weight. A lattice 
configuration is then specified by placing the allowed plaquette
configurations on the shaded squares of the lattice shown in Figure
\ref{fig:ising_lattice}. 
The open plaquettes in this diagram play the role of passive
plaquettes, whose configurations are determined only by the surrounding
configurations shown in Figure \ref{fig:vertices}. Every configuration
is allowed for an open plaquette and each occurs with weight 1.
The partition function for this system is
\begin{equation}
Z =
\sum_{\begin{array}{c}{\rm allowed}\\{\rm
configurations}\end{array}}
\prod_{{\rm plaquettes}\ i} \omega_i \;,
\end{equation}
As mentioned previously, the division of the quantum Hamiltonian into
sums over commuting operators must be chosen with a specific update
mechanism in mind. From Figure \ref{fig:ising_lattice}, we see that
the division (\ref{H0}) to (\ref{H3}) combined with the insertion of
basis states in equation (\ref{insertion}) describes a parallel
sublattice update mechanism  --- the shaded plaquettes in a row of the
diagram represent the bonds chosen for an update in one
time-step.  Alternatively, we may have chosen to
describe sequential update. This is achieved by dividing the
quantum Hamiltonian into $L$ pairs of local bond Hamiltonians. 
One of each pair contains only diagonal terms, the other contains only
non-diagonal terms, and there is one pair for every bond in the system. 
Then, inserting basis states between each pair of local Hamiltonians
leads to a sequential update mechanism.  

We also note that in the chosen basis, we can write a transfer matrix
for the evolution of configurations along the Trotter axis. Its matrix
elements are identical to those appearing in the matrix representation of the
discrete time Master equation (for a specified update mechanism) with
a hopping rate given by $\beta p/n$.
 
\subsection{Scaling}

By rewriting the quantum Hamiltonian as a classical Hamiltonian, the
problem of non-commutation of operators in the quantum Hamiltonian has
been removed and the classical Hamiltonian provides Boltzmann weights
for physical processes (in the sense that each plaquette configuration
has a direct physical interpretation in terms of the original
stochastic dynamics). The restricted number of plaquette
configurations and the way they are placed on the lattice suggest that
we can devise direct and straightforward scaling
procedures. Configurations (and their weights) are built up by piecing
plaquettes together in an allowed way. Then one can coarse-grain by
matching configurations under a change of length scale. This dilation
of scale may be applied to either the time axis or the spatial axis
individually, or to both at once. One can also explore how changing
the update mechanism effects the scaling.

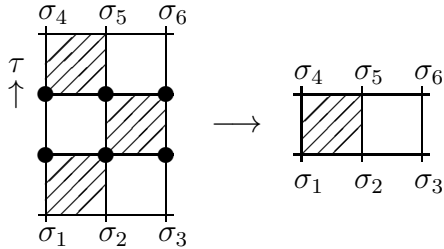
\begin{figure}
\begin{center}
\unitlength=1mm
\linethickness{0.4pt}
\begin{picture}(60,38)

\put(4,5){\line(1,0){18}}
\put(4,13){\line(1,0){18}}
\put(4,21){\line(1,0){18}}
\put(4,29){\line(1,0){18}}
\put(38,13){\line(1,0){18}}
\put(38,21){\line(1,0){18}}

\put(5,4){\line(0,1){26}}
\put(13,4){\line(0,1){26}}
\put(21,4){\line(0,1){26}}
\put(39,12){\line(0,1){10}}
\put(47,12){\line(0,1){10}}
\put(55,12){\line(0,1){10}}

\put(5,5){\line(1,1){8}}
\put(5,7){\line(1,1){6}}
\put(5,9){\line(1,1){4}}
\put(5,11){\line(1,1){2}}
\put(7,5){\line(1,1){6}}
\put(9,5){\line(1,1){4}}
\put(11,5){\line(1,1){2}}
\put(5,21){\line(1,1){8}}
\put(5,23){\line(1,1){6}}
\put(5,25){\line(1,1){4}}
\put(5,27){\line(1,1){2}}
\put(7,21){\line(1,1){6}}
\put(9,21){\line(1,1){4}}
\put(11,21){\line(1,1){2}}
\put(13,13){\line(1,1){8}}
\put(13,15){\line(1,1){6}}
\put(13,17){\line(1,1){4}}
\put(13,19){\line(1,1){2}}
\put(15,13){\line(1,1){6}}
\put(17,13){\line(1,1){4}}
\put(19,13){\line(1,1){2}}

\put(39,13){\line(1,1){8}}
\put(39,15){\line(1,1){6}}
\put(39,17){\line(1,1){4}}
\put(39,19){\line(1,1){2}}
\put(41,13){\line(1,1){6}}
\put(43,13){\line(1,1){4}}
\put(45,13){\line(1,1){2}}

\put(5,13){\circle*{2}}
\put(13,13){\circle*{2}}
\put(21,13){\circle*{2}}
\put(5,21){\circle*{2}}
\put(13,21){\circle*{2}}
\put(21,21){\circle*{2}}

\put(27,16){$\longrightarrow$}
\put(4,1.5){$\sigma_1$}
\put(12,1.5){$\sigma_2$}
\put(20,1.5){$\sigma_3$}
\put(4,31){$\sigma_4$}
\put(12,31){$\sigma_5$}
\put(20,31){$\sigma_6$}

\put(38,8.5){$\sigma_1$}
\put(46,8.5){$\sigma_2$}
\put(54,8.5){$\sigma_3$}
\put(38,23){$\sigma_4$}
\put(46,23){$\sigma_5$}
\put(54,23){$\sigma_6$}

\put(0,20){$\uparrow$}
\put(0,24){$\tau$}

\end{picture}
\caption{Decimation of the Trotter lattice along the time axis. Sites
labelled by a circle are traced over.} \label{fig:ising_dec}
\end{center}
\end{figure}

To illustrate these ideas, consider a putative decimation eliminating
sites along the time axis for a parallel sublattice update
mechanism. This is shown, for a particular matching of active and
inactive plaquettes, in Figure \ref{fig:ising_dec}.
The weight for the unrenormalised system is obtained by tracing over
all allowed plaquettes consistent with a given configuration $\{
\sigma_i\}$. A scaling equation may be obtained, for example, by
matching this weight with the renormalised weight for a plaquette with
spins given by $\sigma_1$, $\sigma_2$, $\sigma_4$ and $\sigma_5$.


\section{Conclusion}

Scaling techniques which are simple to implement have been described
within the quantum formulation of nonequilibrium exclusion models. In
particular, the exact fixed points are obtained for the ASEP in which
the known dependence of the dynamics on asymmetry is recovered. For a
model involving the evaporation and deposition of adjacent pairs
exact fixed points are obtained; the resulting flow diagram indicates
that the dynamics are independent of the ratio of evaporation and
deposition rates. Further, the stability of the
free-fermion condition in the model combining the dynamics of the ASEP
with pair evaporation and deposition indicates that the exact
solutions provide a complete account of the large scale
behaviour. Again, the dynamics are found to be independent of bias.

We note that the projections of spin-1 operators onto a basis of
spin-1/2 operators in Sections \ref{sect:quant_ASEP} and
\ref{sect:quant_PED} are not necessary. Indeed, one could continue
scaling the system to renormalised bases of higher and higher
spin. An approach to yield the continuum limit of the ASEP in this way
\cite{fogedby95} is under investigation. 

We have also shown how to map a quantum Hamiltonian representing a
nonequilibrium exclusion process onto a classical Hamiltonian in one
higher dimension. In this way, we have shown how to write the steady
state probabilities for configurations of the nonequilibrium system in
terms of classical Boltzmann weights. This should enable one to borrow
real-space renormalisation group techniques for classical Hamiltonian
systems and apply them to nonequilibrium systems.

\end{document}